\documentclass[%
preprint,
superscriptaddress,
 amsmath,amssymb,
 aps,
]{revtex4-1}

\usepackage{graphicx}
\usepackage{dcolumn}
\usepackage{bm}
\usepackage{amsmath}
\usepackage{tikz,comment}

\usepackage{natbib}
\usepackage{graphicx}
\usepackage{epstopdf,epsfig}
\usepackage{mathrsfs}
\usepackage{graphicx,amssymb,amsmath,color}
\usepackage{soul}
\usepackage{mathtools}

\begin{document}

\preprint{APS/123-QED}

\title{Bistability in the wake of a wavy cylinder}

\author{Kai Zhang}\thanks{Current address: Department of Mechanical and Aerospace Engineering, University of California, Los Angeles. kzhang3@ucla.edu}

\affiliation{School of Naval Architecture, Ocean \& Civil Engineering, Shanghai Jiao Tong University, Shanghai 200240, China}

\author{Dai Zhou}\thanks{Corresponding author: zhoudai@sjtu.edu.cn}
\affiliation{School of Naval Architecture, Ocean \& Civil Engineering, Shanghai Jiao Tong University, Shanghai 200240, China}
\affiliation{Key Laboratory of Hydrodynamics of Ministry of Education, Shanghai, 200240, China}
\affiliation{State Key Laboratory of Ocean Engineering, Shanghai Jiao Tong University, Shanghai, 200240, China}

\author{Hiroshi Katsuchi}
\affiliation{Department of Civil Engineering, Yokohama National University, Yokohama, Kanagawa 2408501, Japan }

\author{Hitoshi Yamada}
\affiliation{Department of Civil Engineering, Yokohama National University, Yokohama, Kanagawa 2408501, Japan }

\author{Zhaolong Han}
\affiliation{School of Naval Architecture, Ocean \& Civil Engineering, Shanghai Jiao Tong University, Shanghai 200240, China}

\author{Yan Bao}
\affiliation{School of Naval Architecture, Ocean \& Civil Engineering, Shanghai Jiao Tong University, Shanghai 200240, China}

\date{\today}

\begin{abstract}

We investigate the wake dynamics of an optimally designed wavy cylinder that completely suppresses the K\'arm\'an vortex shedding. Such a wavy cylinder is forced to oscillate with a sinusoidal motion in the crossflow direction. 
Examination of the lift force spectrum reveals that a critical forcing frequency exists, below which the flow control effectiveness of the wavy cylinder is retained, and beyond which the inherent vortex shedding resurrects. Moreover, the resurrected unsteady vortex shedding can persist even without sustained forcing, leading to the loss of control efficacy. This suggests that in addition to the steady state developed from uniform initial condition, an oscillatory state exists in the wake of wavy cylinder if the initial state is sufficiently perturbed. The discovery of the bistable states calls for examination of the flow control effectiveness of the wavy cylinder in more complicated inflow conditions.

\end{abstract}

\maketitle


\section{Introduction}
\label{sec:intro}
Three-dimensional forcing techniques, which apply spanwise varying controls along nominally two-dimensional bluff body, have been recognized effective in controlling the vortex shedding in the wake \citep{choi2008control}. Among the various realizations of this class of methods, circular cylinder with spanwise varying diameter (referred to as wavy cylinder hereafter) has attracted a lot of attention due to its omnidirectional shape. 
Early experimental works by \citet{ahmed1992transverse,ahmed1993experimental} have revealed that pressure gradient exists in the spanwise direction, leading to the formation of non-uniform separation lines along the span and the development of the three-dimensional wake. The subsequent investigations discovered that such three-dimensional wake is associated with significant suppression of the K\'arm\'an vortex shedding and reduction in the drag and lift forces \citep{lam2004experimental,lam2009effects,xu2010large,lam2008large,lin2016effects}. Moreover, the spanwise-undulated geometry is found to resemble the whiskers of harbor seals. Such particular shape has been shown to exhibit superior hydrodynamic performance and to enhance the sensitivity of the whisker even in turbid water \citep{hanke2010harbor}. This finding has inspired the bio-mimicry innovations like the energy conserving flow sensors \citep{beem2012calibration}. 

The flow control effectiveness of the wavy cylinder is originated from the streamwise vortical structures in the near wake, as evidenced by a number of experimental and computational works \citep{zhang2005piv,lam2009effects,zhang2018large}.
Embedded in the three-dimensional free shear layers, these vortical structures appear as counter-rotating pairs within each wavelength. The existence of such streamwise structures in the near wake tends to inhibit the roll-up of the vortex sheets along the spanwise direction, thus delaying the formation of the K\'arm\'an vortex shedding. With optimal shape parameters, the wake unsteadiness could even be completely concealed \citet{lam2009effects}. On the theoretical side, \citet{hwang2013stabilization} conducted linear stability analysis of the spanwise-wavy wake profiles. Using Floquet theory, they found that the introduction of the spanwise waviness attenuates the absolute instability of the two-dimensional wakes. 

Most of the previous studies on the wavy cylinders have focused on the simple configuration of uniform flow over a fixed body. However, configurations of practical interest exist, notably in bridge cables and deepwater risers, in which the slender cylindrical structures could be subjected to vibrations that are either self-induced or externally forced. Recently, \citet{zhang2017numerical} conducted numerical investigation for the vortex-induced vibration (VIV) of the wavy cylinder. Despite the complete suppression of vortex shedding in the fixed configuration, the wavy cylinder is able to develop large-amplitude crossflow vibrations when flexibly mounted, and the vibration response curve is similar to that of a two-dimensional circular cylinder at the same conditions. The numerical results were verified by \citet{assi2018vortex}, who conducted water tunnel experiment on the VIV of elliptical wavy cylinders. They have shown that once the cylinders start to oscillate, the separation lines straighten up, and the spanwise-coherent vortex filaments dominate the near wake, recovering a wide K\'arm\'an wake.

The distinct behaviors of the wavy cylinder in the fixed configuration and in motion entail further examination of its flow control effectiveness, which is attempted in the current paper using direct numerical simulations.
The wavy cylinder is optimally designed so that it completely suppresses the wake unsteadiness at a range of Reynolds numbers. To investigate the wake dynamics such a wavy cylinder, we force the cylinder to oscillate in the crossflow direction with prescribed motion.
The behavior of the force coefficients in the perturbed flow, as well as the development of the wake, is analyzed. The current numerical results reveals the existence of the bistable state in the wake of the wavy cylinder, i.e., a steady state (fixed point) if the cylinder is slightly perturbed, and an oscillatory state (limited cycle) in highly perturbed flow.
The rest of the paper is organized as follows. In \S \ref{sec:setup}, we present the problem description and numerical setup. In \S \ref{sec:static}, the flow over the fixed wavy cylinder is investigated to locate the control-effective Reynolds numbers. Then, we focus on the wake dynamics of the oscillating wavy cylinder in \S \ref{sec:forced}. Further in \S \ref{sec:bistability}, we prove the existence of the bistability by letting the wake develop from different initial conditions. Finally, we conclude this paper by summarizing our findings in \S \ref{sec:conclusion}.

\section{Computational setup}
\label{sec:setup}

\begin{figure}
	\centering
	\includegraphics[scale=0.55]{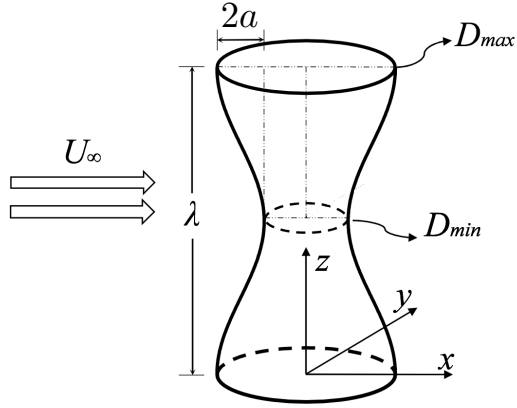}
	\caption{Problem description and coordinate system.}
	\label{fig:scheme}
\end{figure}

The geometry of the wavy cylinder is schematically depicted in figure \ref{fig:scheme}. The diameter of the wavy cylinder varies sinusoidally along the spanwise direction $z$ according to
\begin{equation}
D(z)=D_m+2a\cos(2\pi z/\lambda),
\end{equation}
where $D_m$ is the averaged diameter, $a$ and $\lambda$ are the geometric amplitude and wavelength, respectively. In the current paper, we assign $a=0.175D_m$ and $\lambda=2.5D_m$. This set of parameters has been shown by \citet{lam2009effects} to exhibit satisfactory flow control efficacy. The wavy cylinder is subjected to a uniform incoming flow $U_{\infty}$ in the $x$ direction. For non-dimensionalization, we normalize the spatial variables by the averaged diameter $D_m$, velocity by $U_{\infty}$, time by $D_m/U_{\infty}$, and frequency by $U_{\infty}/D_m$. The Reynolds number, defined as $Re\equiv U_{\infty}D_m/\nu$, where $\nu$ is the kinematic viscosity of the fluid, is kept below 160.

The wavy cylinder is forced to vibrate in the crossflow ($y$) direction with the prescribed motion
\begin{equation}
    y(t)=A_e\sin(2\pi f_e t),
\end{equation}
in which $A_e$ is the nondimensional forcing amplitude, and $f_e$ is the dimensionless forcing frequency. 

The flow is governed by the incompressible Navier-Stokes equations, which are solved by the direct numerical simulations using the open-source software OpenFOAM. Both the time and space are discretized with second-order accurate schemes. The wavy cylinder is placed in the center of a circular computational domain of $30D_m$ in radius and $2.5D_m$ in height. Periodic boundary conditions is specified at the spanwise ends of the domain. The cylinder surface is treated as no-slip wall. A uniform flow condition with freestream velocity $U_{\infty}$ is specified at the inlet. 

An O-type mesh with resolution $N_c\times N_r\times N_z=140\times 140 \times 40$ (where $N_c$, $N_r$ and $N_z$ represent the grids in the circumferential, radial and spanwise directions) is used for the domain discretization. The mesh is concentrated in the vicinity of the cylinder to better resolve the near wake. The nondimensional time-step is set to be $\Delta t=0.02$. The convergence of the numerical results against the grid resolution has been verified through a mesh dependency test, as shown in table \ref{tab:kd}. The mean drag coefficient $\overline{C_D}$ and the root-mean-squred lift coefficient $C_L^{\prime}$ are reported for three cases: flows over fixed wavy cylinders at $Re=100$ and $Re=150$, and flow over an oscillating wavy cylinder with $(A_e,f_e)=(0.2,0.28)$ at $Re=150$. Here, the drag and lift coefficients are defined as $C_D=2F_D/(\rho U_{\infty}^2D_mH)$ and $C_L=2F_L/(\rho U_{\infty}^2D_mH)$, in which $F_D$ and $F_L$ are the drag and lift forces, and $\rho$ the fluid density. It is observed that when the grid resolution is increased to \#2, the aerodynamic forces become converged upon further refinement of the mesh. Besides, the drag and lift coefficients of the fixed wavy cylinder at $Re=100$ are in agreement with the values reported in \citet{lam2009effects} for the same geometry (extracted from figure 4 of their paper). Thus, the mesh resolution \#2 is used throughout this paper.

\begin{table*}
	\begin{center}
\begin{tabular}{p{0.8cm}p{1cm}p{1cm}p{1cm}p{1.2cm}p{0.5cm}p{1cm}p{1cm}p{1cm}p{0.5cm}p{1cm}p{1cm}p{1cm}}
	& \multicolumn{4}{c}{$Re=100$, fixed} & & \multicolumn{3}{l}{$Re=150$, fixed} &    & \multicolumn{3}{l}{$Re=150$, oscillating} \\ \hline
    & 	 \#1 & \#2 & \#3  & Ref \citep{lam2009effects} & & \#1 & \#2 & \#3  & &   \#1 &   \#2 &   \#3 \\
	$\overline{C_D}$   &  1.33   &  1.35   &   1.35  &  1.35  &   &     1.00 &     1.01 &    1.01  &  & 1.33   &    1.35 &      1.36   \\
	$C_L^{\prime}$  &  0.20  &  0.21   &   0.21   &   0.21 &   &  0    &   0  &  0 &   &1.05   &  1.09   &  1.09 \\
\end{tabular}
		\caption{Mesh dependency test. The mesh resolution in case \#1 is $N_c\times N_r \times N_z=120\times 120\times 30$, case \#2: $N_c\times N_r \times N_z=140\times 140\times 40$, and case \#3: $N_c\times N_r \times N_z=160\times 160\times 60$. For the oscillating cylinder case at $Re=150$, $(A_e,f_e)=(0.2,0.28)$.}{\label{tab:kd}}
	\end{center}
\end{table*}

\section{Static configuration}
\label{sec:static}

\begin{figure}
	\centering
\includegraphics[scale=0.54]{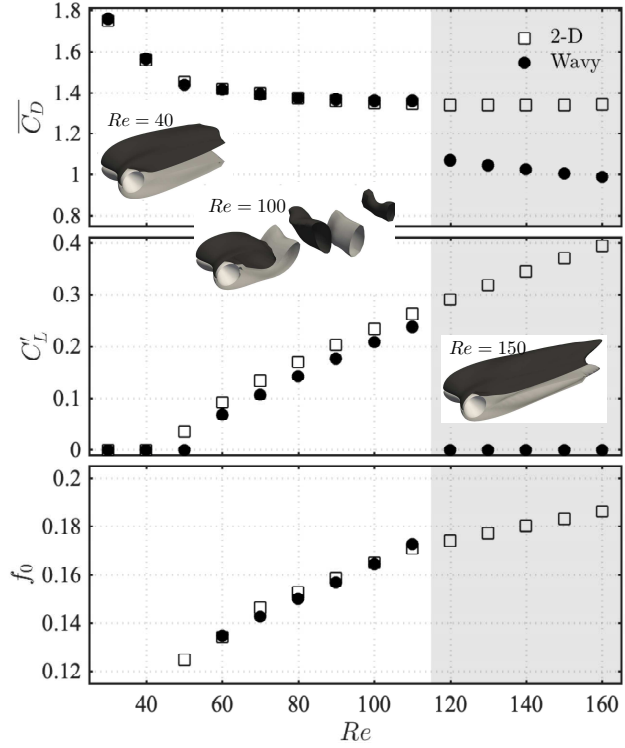}
	\caption{Mean drag, rms lift coefficients and the shedding frequency of flow past two-dimensional and wavy cylinders in the fixed configuration. Iso-surfaces of $\omega_z=-0.5$ (dark color) and $0.5$ (light color) at $Re=40,$ 100 and 150 are included. Shaded area indicates the control-effective regime.}
	\label{fig:static}
\end{figure}

The flow around the static wavy cylinder subjected to uniform initial condition is examined at $Re=30-160$ to locate the control-effective regime. The mean drag coefficient $\overline{C_D}$, root-mean-squared lift coefficient $C_L^{\prime}$ and nondimensional shedding frequency $f_0$ are shown in figure \ref{fig:static}. For $Re\lesssim 110$, the wakes of both the wavy and two-dimensional cylinders transit from steady state at low $Re$ to vortex shedding at higher $Re$, although the critical Reynolds number of the transition is slightly larger for the wavy cylinder. The drag, lift and Strouhal number of the two cylinders are also similar to each other. As the Reynolds number is increased to 120, great flow control efficacy is achieved by the wavy cylinder. The drag force suffers from a drastic decrease compared with the two-dimensional cylinder. The lift force drops to zero, suggesting the complete suppression of vortex shedding in the wake of the wavy cylinder. The steady flow achieved at $Re\gtrsim 120$ is due to the the formation of streamwise vortical structures in the near wake inhibiting the roll-up of the free shear layers, as elucidated in \citet{lam2009effects,hwang2013stabilization}.

\section{Forced vibration}
\label{sec:forced}

\begin{figure}
	\centering
	\includegraphics[scale=0.54]{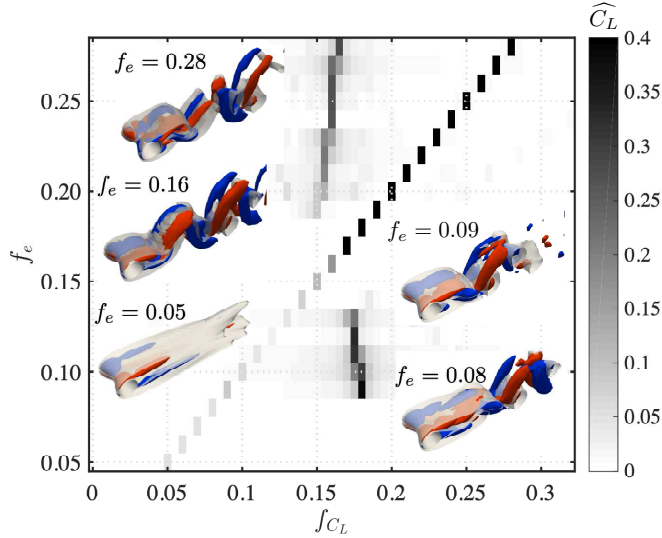}
	\caption{Lift spectrum of the wavy cylinder undergoing forced vibration with $A_e=0.2$ at $Re=150$.  Instantaneous vortical structures are presented at several forcing frequencies, with transparent gray standing for iso-surface of $\omega_z=\pm 0.5$, and red and blue for $\omega_x=0.3$ and -0.3, respectively.}
	\label{fig:wavyForced}
\end{figure}

Now that the control-effective regime of the wavy cylinder is located, let us perturb the cylinder with sinusoidal oscillations in the crossflow direction and investigate its wake dynamics. The lift force spectrum of the oscillating wavy cylinder at $Re=150$ is shown in figure \ref{fig:wavyForced}(\textit{a}). The amplitude of forcing is fixed at $A=0.2$ and the forcing frequency $f_e$ is varied from $0.05$ to $0.28$, at an interval of $0.01$. The vortical structures are visualized by iso-surfaces of $\omega_z$ and $\omega_x$ for selected cases. With small forcing frequency $f_e=0.05$, the lift coefficient is dominated by a single frequency at $f_e$. The vortical structures in the wake sway slightly with the motion of the cylinder. The streamwise vorticity remain in the near wake and play its role in suppressing the roll-up of the free shear layer. The vortical structure at $f_e=0.08$ appears more unsteady. Nevertheless, still a single peak at $f_e$ is observed in the lift spectrum. The situation is significantly different when it comes to $f_e \geq 0.09$. In the lift spectrum, apart from the forcing frequency, another peak reminiscent of the inherent shedding frequency as in the two dimensional case \citep{carberry2005controlled,kumar2016lock} emerges at the nondimensional frequency of $f_s =0.15\sim 0.17$. This is also manifested in the corresponding wake vortical structures, where the roll-up of the free shear layers occurs much closer to the cylinder compared with $f_e=0.08$. Along with the spanwise vortical structures, the periodic shedding of the streamwise vorticity is also observed. With the revival of the inherent vortex shedding, the wake of the wavy cylinder is able to lock onto the forcing at $f_e=0.14\sim0.18$. Further increasing the forcing frequency reveals $f_s$ again, although its value is slightly smaller than that at smaller forcing frequencies. It has been confirmed that the resurrection of the inherent shedding frequency of a forced oscillating wavy cylinder also occurs at much higher Reynolds number of $Re=5000$ \citep{zhang2018numerical}.

\begin{figure}
\centering
\includegraphics[scale=0.55]{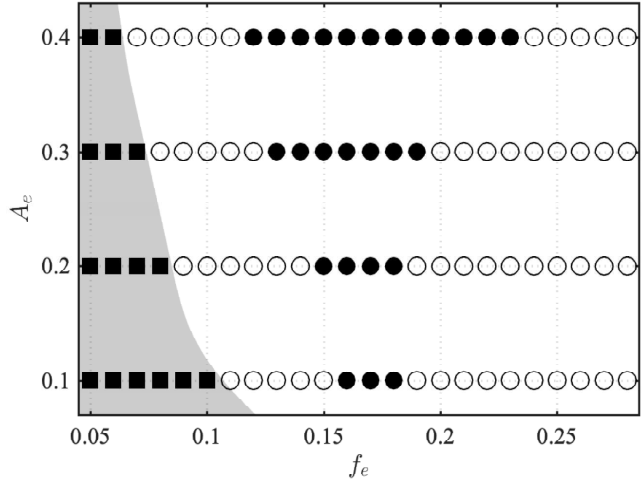}
\caption{Classification of wake states of the wavy cylinder in forced vibration at $Re=150$. Solid square in shaded region: control-effective. Unshaded region: control effect lost. Solid circle: lock-on. Empty circle: lock-out. }
\label{fig:regime}
\end{figure}

We also test the cases with different forcing amplitudes $A_e$ with varying forcing frequencies $f_e$. Based on the lift spectra of these cases, a classification of the wake states of the wavy cylinder undergoing forced vibration is presented in figure \ref{fig:regime}. The wavy cylinder maintains its flow control effectiveness when the forcing frequency $f_e$ is smaller than a critical value. This critical value  decreases with increasing forcing amplitude $A_e$. In this regime, only a single peak is found in the lift spectrum. As the forcing frequency exceeds the critical value, the inherent shedding frequency $f_s$ appears in the lift spectrum, siding with the forcing frequency $f_e$. Similar to a two-dimensional cylinder, the inherent vortex shedding can submit to the forcing when the two frequencies are close, giving rise to the lock-on region that features an Arnold tongue in the $f_e$-$A_e$ space.

\section{Bistability in the wake}
\label{sec:bistability}
\begin{figure}
	\centering
\includegraphics[scale=0.56]{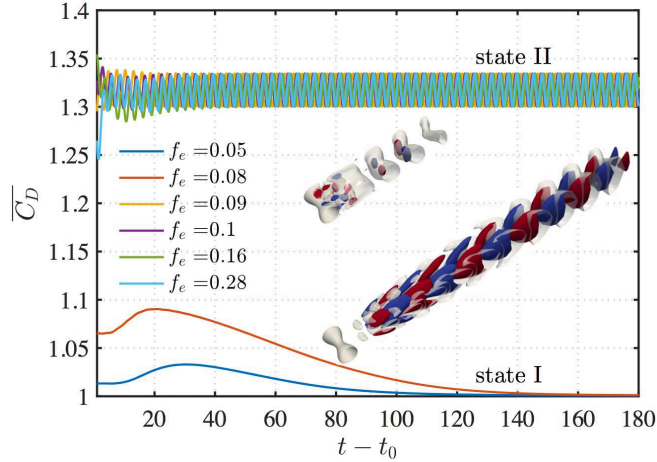}
	\caption{Dependence of drag coefficients on the initial conditions at $Re=150$. The time at which the forced vibration is stopped is denoted as $t_0$. The bottom inset is the linear stability mode for the steady state I. The top inset is the DMD mode corresponding to the primary frequency of the wake. Transparent gray represents the iso-surfaces of $\tilde{\omega}_z=\pm 0.005$. Red and blue represents iso-surfaces of positive and negative $\tilde{\omega}_x$ of the same contour level. }
\label{fig:initial}
\end{figure}

More interestingly, with the resurrection of inherent shedding frequency $f_s$, the unsteady vortex shedding is able to persist by itself even without sustained forcing. To prove this, we manually turn off the forced vibration when the cylinder reaches the top or bottom position, at which the velocity of the cylinder becomes zero. This ensures a smooth transition from the dynamic simulation to a static one. We then focus on the evolution of the wake starting from the initial condition dictated by forced vibration. The time histories of $C_D$ obtained by this procedure are presented in figure \ref{fig:initial}(\textit{a}) for $A_e=0.2$ with selected forcing frequencies. For cases with $f_e\lesssim 0.09$, the drag coefficients exhibit slight increase over time but eventually converge to a fixed value of $C_D=1.0$ (state I) as reported in \S \ref{sec:static}. On the other hand, at $f_e \gtrsim 0.09$, for which the inherent shedding in the forced wake has revived, the flow eventually arrives at state II with the drag coefficient oscillating at around $\overline{C_D}= 1.32$, signifying the loss of flow control efficacy. We further perform dynamic mode decomposition (DMD) to extract the coherent structures at both states. In the case of steady state I, DMD is conducted on snapshots in the linearly decaying regime, thus the modal structure, as shown in the bottom inset, is identical to the linear stability mode \citep{ferrer2014low,tu2014dynamic,theofilis2011global}. This mode features high three dimensionality with prominent streamwise strucutres. The DMD mode for the oscillatory state II is conducted on the periodical shedding regime and is plotted in the upper inset. In this mode, the streamwise structures that are prominent in the linear stability mode appear to be less active. This modal structure is similar to that of the fixed wavy cylinder at $Re\lesssim 110$. The comparison of the modal structures for the two states highlights the importance of streamwise vortical structures in maintaining the wake stability.

\begin{figure}
	\centering
\includegraphics[scale=0.58]{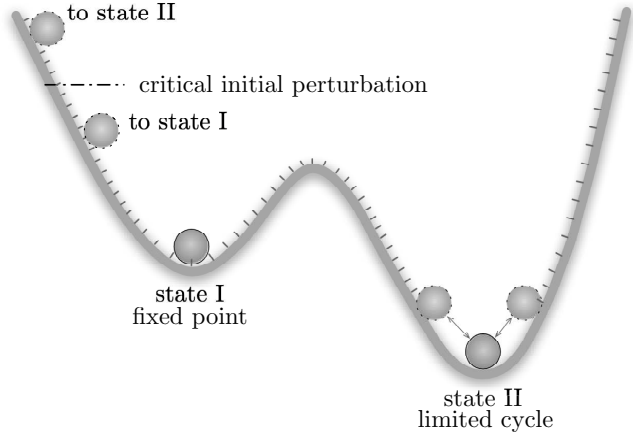}
	\caption{Schematic diagram of the bi-stable states. State I represents the steady wake and state II the oscillatory wake.}
\label{fig:bistable}
\end{figure}

A schematic diagram for the bistable states of the wavy cylinder wake is shown in figure \ref{fig:bistable}. Both the steady state I and oscillatory state II are stable so that there exists a barrier between the two states. For weakly disturbed flow, the oscillations in the flow are damped and the wake eventually converges to the steady state I. However, the barrier is easily overcome when the initial condition is sufficiently perturbed, causing the flow to overshoot to the oscillatory state II. Such strong dependency of the long-term flow states on the initial condition is in glaring contrast to the conventional two-dimensional bluff body flows, for which the effect of initial condition is usually limited in time, and the long-term state depends only on the Reynolds number \citep{laroussi2014triggering}. 

The bistability in the wake of the wavy cylinder is a result of the competition between the streamwise vortical structures that attempts to stabilize the flow \citep{lam2009effects,hwang2013stabilization}, and the absolute instability that tends to destabilize the wake \citep{zebib1987stability,jackson1987finite}. While the latter mechanism is ever-present at super-critical Reynolds numbers, the streamwise vortices are susceptible to external disturbances. Once the inherent shedding is triggered, say, by structural oscillation, the steady streamwise vortical structures that are responsible for the wake stabilization are compelled to oscillate by the spanwise vortices and could no longer return to its initial state. As a result, the wake surrenders to the periodic K\'arm\'an vortex shedding and the flow control efficacy is lost. 

We note in passing that the existence of the oscillatory state is not the direct cause for the onset of vortex-induced vibration (VIV) of the wavy cylinders in uniform incoming flows \citep{zhang2017numerical,assi2018vortex}. Instead, the destabilization of the wavy cylinder from steady flow is comparable to the VIV of a two-dimensional cylinder at subcritical Reynolds numbers of $Re\lesssim 47$ \citep{mittal2005vortex,kou2017lowest}. Recent works have revealed that VIV occurs from the linear instability of the coupled fluid-structure system \citep{zhang2015mechanism,mittal2016lock,yao2017model}. On the other hand, the emergence of the oscillatory state in the wavy cylinder wake requires considerable forcing that is beyond the linear assumption. Once the flow is sufficiently perturbed by the vibration, the oscillatory state could be triggered and interact with the motion of the cylinder. 

\section{Conclusion}
\label{sec:conclusion}
Direct numerical simulations have been conducted to study the wake dynamics of a wavy cylinder at low Reynolds numbers. The wavy cylinder is optimally designed so that it completely suppresses wake unsteadiness in the fixed configuration. Deeper insights are obtained by perturbing the flow with sinusoidal structural oscillations with varying frequencies. It is disclosed that the control efficacy of the wavy cylinder could only be preserved with weak forcing. As the forcing frequency exceeds a critical value, the inherent shedding frequency that has been concealed in the fixed configuration revives, further leading the flow to lock-in. The resurrected inherent shedding vortices could persist even without sustained forcing, implying the existence of the bistable states in the wavy cylinder wake. 

In realistic applications, the transition from the steady state to the oscillatory state in the wavy cylinder wake could be triggered in many scenarios such as forced vibration, self-induced vibration, highly turbulent incoming flow, gusty winds, just to name a few. The discovery unveiled from this work calls for reexamination of the control effectiveness of the wavy cylinder in more complicated flow conditions.

\begin{acknowledgments}
The financial support from the National Natural Science Foundation of China (Nos. 51679139,11772193and 51879160), the Innovation Program of Shanghai Municipal Education Commission (No.2019-01-07-00-02-E00066) ，the Shanghai Natural Science Foundation (No. 17ZR1415100 and 18ZR1418000) and the Program for Intergovernmental International S\&T Cooperation Projects of Shanghai Municipality (No. 18290710600), are gratefully acknowledged. This research was also supported in part by the Program for Professor of Special Appointment (Eastern Scholar) at Shanghai Institutions of Higher Learning (No. ZXDF010037), the Project of Thousand Youth Talents (No. BE0100002) and the Major Program of the National Natural Science Foundation of China
(No. 51490674). KZ thanks Prof. Kunihiko Taira and Dr. Chi-An Yeh for their useful comments. 
\end{acknowledgments}

\bibliography{refs.bib}

\end{document}